\documentclass[11pt,a4paper]{article}
\usepackage{amsmath, amsthm, amssymb} 
\usepackage{amsfonts} 
\usepackage{latexsym}
\usepackage{amsmath,amscd}
\usepackage{graphicx}
\usepackage[margin=1in]{geometry}
\usepackage{makeidx}
\usepackage[english]{babel}
\usepackage{changepage}  
\usepackage{color}
\usepackage{pict2e}
\usepackage{hyperref}

\DeclareMathOperator{\coker}{coker}

\DeclareMathOperator{\image}{Im}
\DeclareMathOperator{\real}{Re}

\newcommand*\xbar[1]{%
  \hbox{%
    \vbox{%
      \hrule height 0.5pt 
      \kern0.5ex
      \hbox{%
        \kern-0.1em
        \ensuremath{#1}%
        \kern-0.1em
      }%
    }%
  }%
}

\def\hybrid{\topmargin -10pt    \oddsidemargin 0pt
        \headheight 0pt \headsep 0pt
       \textwidth 6.25in       
      \textheight 9.5in       
        \marginparwidth .875in
        \parskip 5pt plus 1pt   \jot = 1.5ex}
\hybrid
\theoremstyle{remark}

\theoremstyle{definition}

\begin{document}

\thispagestyle{empty}

\rightline{\small MI-TH-1630}

\vskip 3cm
\noindent
{\LARGE \bf  Defects and boundary RG flows in $\mathbb{C}/\mathbb{Z}_d$}\\
\vskip .8cm
\begin{center}
\linethickness{.06cm}
\line(1,0){447}
\end{center}
\vskip .8cm
\noindent
{\large \bf Melanie Becker$^\dagger$, Yaniel Cabrera$^\dagger$, Daniel Robbins$^\ddagger$}

\vskip 0.2cm

\begin{tabular}{p{3.5in}p{3.5in}}
\hskip -0.15cm ${}^\dagger${\em    George and Cynthia Mitchell Institute} & \hskip -0.15cm ${}^\ddagger${\em Department of Physics} \\
{\em for Fundamental Physics and Astronomy} & {\em University at Albany} \\
{\em Texas A \&M University} & {\em 1400 Washington Ave.} \\
{\em College Station, TX 77843--4242, USA} & {\em Albany, NY 12222, USA}
\end{tabular}

\vskip 0.5cm
{\tt \hskip -.5cm mbecker, cabrera AT physics.tamu.edu, dgrobbins AT albany.edu}

\vskip 1cm

\vskip0.6cm

\noindent
{\sc Abstract:}  We show that topological defects in the language of Landau-Ginzburg models carry information about the RG flow between the non-compact orbifolds $\mathbb{C}/\mathbb{Z}_d$. We show that such defects correctly implement the bulk-induced RG flow on the boundary.

\pagebreak

\newpage

\tableofcontents

\section{Introduction}\label{intro}

The behavior of boundary degrees of freedom under renormalization group (RG) flow represents a problem in both string theory and condensed matter physics that is not fully understood (see \cite{dorey00}, \cite{keller07}, \cite{hori04}, \cite{fredenhagen06} and references therein).  A new approach consists of utilizing defects to bring the RG flow from the bulk to the boundary. A defect is a one-dimensional object in two-dimensional theories, and more generally a codimension-one submanifold in higher dimensional spaces. This technique was exploited in \cite{brunner07a} within the framework of Landau-Ginzburg models to study the boundary RG flow between the two-dimensional orbifolds $\mathcal{M}_{d-2}/\mathbb{Z}_d$, where $\mathcal{M}_{d-2}$ are the supersymmetric minimal models.  RG flow defects were also constructed in \cite{gaiotto12} between consecutive Virasoro minimal models in two dimensions.

 Defects are not restricted to Landau-Ginzburg (LG) models but in these theories they have a general description in terms of matrix factorizations which allows us to construct examples of boundaries and defects. Also, the language of matrix factorization provides a general operation called the tensor product of matrix factorizations which gives a recipe to compute the fusion of any two LG defects \cite{enger05, brunner07, brunner07a}. The theory of defects in Landau-Ginzburg models is versatile because it provides direct information on other theories which are not necessarily LG models. This fact follows because Landau-Ginzburg models can be mapped to other interesting theories via different RG flows or mirror symmetry \cite{hori00, vafa01}. In this article we are particularly interested in the non-compact orbifold $\mathbb{C}/\mathbb{Z}_d$. This orbifold is not target-space supersymmetric, but it exhibits $N=2$ worldsheet supersymmetry. 

The study of defects has been  mainly restricted to two-dimensional conformal field theories. There, the objects of interest are called \emph{conformal defects} which commute with the difference of the holomorphic and antiholomorphic components of the energy-momentum tensor \cite{fuchs07}. A subset are those defects called \emph{topological} which fully commute with the energy-momentum tensor. In this case, the defect can be translated and deformed through the worldsheet without affecting the values of the correlation functions, as long as it does not cross an operator insertion point.

In Landau-Ginzburg models defects are topological provided that a topological twist has been performed \cite{brunner07}. There are two types of twists that render $N=2$ theories topological \cite{hori00a}, and they are called A-twist and B-twist. In the presence of boundaries or defects, only half of the total $(2,2)$ supersymmetry is preserved, and just like for the topological twist there are two ways to break half of the supersymmetry. The remaining symmetry is called A-type or B-type depending on which supersymmetric charges are kept. The topological A(B)-twist is compatible only with A(B)-type supersymmetry if the boundaries and defects are to be supersymmetric and topological. In this note we assume that the Landau-Ginzburg models are already topological. In each case, there is a BRST-operator $Q_A$ or $Q_B$ which characterizes the physical degrees of freedom at the boundary.

The machinery of matrix factorizations for defects can be applied to  theories such as the non-compact case $\mathbb{C}/\mathbb{Z}_d$ which is the archetype for string theory on
\begin{equation}\label{arch}
\mathbb{R}^{d-1,1}\times\mathbb{R}^{10-d}/G ,
\end{equation}
where $G$ is some discrete $SO(d-10)$ subgroup \cite{harvey01}. This important model is linked to the Landau-Ginzburg language in two ways that are exploited in this note. First, by introducing superspace variables the fermionic string theory on $\mathbb{C}/\mathbb{Z}_d$ can be viewed as the orbifold of a LG model with zero superpotential. And second, we can also go from the $\mathbb{C}/\mathbb{Z}_d$ theory to a twisted LG model using mirror symmetry as given in \cite{vafa01}.

In this note we extend the work of \cite{brunner07a} which describes the boundary RG flow in Landau-Ginzburg models and supersymmetric minimal models in terms of topological defects. Our work generalizes the results of \cite{brunner07a} to the non-supersymmetric case of the non-compact $\mathbb{C}/\mathbb{Z}_n$ theories. The orbifold $\mathbb{C}/\mathbb{Z}_d$ is physically relevant because it is the simplest model to study  tachyon condensation \cite{adams01}; in (\ref{arch}), the tachyons are closed strings localized at the fixed points of the orbifold group action.  Techniques to study the RG flow in these models have been considered in \cite{vafa01, harvey01}.

To study the problem at hand, we consider the $\mathbb{C}/\mathbb{Z}_d$ orbifold theory on the upper-half plane $\Sigma = \left\{(x^0,x^1)\in \mathbb{R}^2 \ | \ x^0 \geq 0\right\}$ with B-type supersymmetry. Inserting the identity defect at $x^0=y>0$, we can perturb the theory over $x\geq y$. Letting the perturbations drive the theory to the IR we obtain a setup describing the IR theory in the bulk while near the boundary we still have the UV theory, with a defect $D$ sitting at the interface $x^0=y$. The next step is to take the RG flow to the boundary via the limit $y\rightarrow 0$. In terms of defect language, this limit gives the fusion of the boundary $B$ and the defect $D$.

The paper is structured as follows. In Section 2 we review $\mathcal{N}=(2,2)$ theories in the presence of boundaries. The introduction of a boundary reduces the supersymmetry and we are left with either A-type or B-type supersymmetry which are halves of the full $\mathcal{N}=(2,2)$ symmetry. We review the algebraic language of matrix factorizations suitable fo B-type boundaries and defects and the geometrical description of wave-front trajectories for A-type boundaries.

Section 3 contains the superspace description of $\mathbb{C}/\mathbb{Z}_d$ as a LG model with zero superpotential. Here we have a description of boundary conditions and defects in terms of matrix factorizations of $W(X)=0$. We show that suitable defects exist such that they divide the UV and IR theories. In the case of $\mathbb{C}/\mathbb{Z}_d$ we can keep track of both RG endpoints by means of the chiral ring. By adding terms to the Lagrangian which induce the RG flow the chiral ring is deformed as well.  The resulting chiral ring at each endpoint of the flow characterizes the theory in the UV or IR.

Lastly, in Section 4 we show that LG defects can be used to work out the boundary RG flows of these theories. We work with the mirror theories of the non-compact orbifolds which are orbifolded LG theories with non-zero superpotentials. The B-type boundary conditions have a dual description in terms of A-branes. We compare the action of the special B-type defects on the B-type boundaries with the RG flow as described by the dual A-type branes. This comparison indicates that indeed the special defects enforce the RG flow on the boundary without a need for regularization techniques.

\section{Landau-Ginzburg models with boundaries}

We work in two dimensions with $\mathcal{N}=(2,2)$ supersymmetry. The general supersymmetric variation is infinitesimally
\begin{equation}\label{fullvar}
\delta_{\epsilon,\bar \epsilon}=\epsilon_+Q_ - -\epsilon_-Q_+ -\bar \epsilon_+\xbar Q_-+\bar\epsilon_-\xbar Q_+,
\end{equation}
where the operators $\left\{Q_+, \xbar Q_+; Q_-,\xbar Q_-\right\}$ are represented by the differential operators

\begin{equation}\label{qdefinition}
Q_\pm =\frac{\partial}{\partial \theta^\pm}+i\bar \theta^\pm\partial_\pm\ ,\ \ \ \ \xbar Q_\pm =-\frac{\partial}{\partial \bar \theta^\pm}-i \theta^\pm\partial_\pm \ ,
\end{equation}
where $\partial_\pm = \frac{\partial}{\partial x^\pm}:= \frac{1}{2}\left(\frac{\partial}{\partial x^0} \pm \frac{\partial}{\partial x^1}\right)$. These supersymmetry generators obey the algebra
\begin{equation}
\left\{Q_+,\xbar Q_+\right\}=-2i \partial_+ \ , \ \ \ \left\{Q_-,\xbar Q_-\right\}=-2i \partial_-\ ,
\end{equation}
with all other anticommutators zero.

In superspace, Landau-Ginzburg models are supersymmetric theories described by the action $S= S_D+S_F$ with
\begin{equation}\label{fterm}
S_D= \displaystyle \int d^2x d^4\theta K(X_i,\xbar X_i),
\end{equation}
\begin{equation}\label{fterm}
S_F= \displaystyle \int d^2x d^2\theta \ W(X_i)\big |_{\bar \theta^{\pm}=0}+\displaystyle \int d^2x d^2\bar \theta \ \xbar W(\xbar X_i)\big |_{\theta^{\pm}=0} \ .
\end{equation}
The action is a functional of superfields $X_i$ which are chiral, i.e. $\xbar D_\pm X_i= 0$ where $\xbar D_\pm$ is the anti-chiral half of the supersymmetric covariant derivative. These operators are defined as
\begin{equation}\label{ddefinition}
D_\pm =\frac{\partial}{\partial \theta^\pm}-i\bar \theta^\pm\partial_\pm\ ,\ \ \ \ \xbar D_\pm =-\frac{\partial}{\partial \bar \theta^\pm}+i \theta^\pm\partial_\pm.
\end{equation}

The smooth function $K$ is called the K\"ahler potential and the holomorphic function $W$ the superpotential. This action is explicitly invariant under supersymmetry when the worldsheet is $\mathbb{R}^2$ or an open subset of it. In this paper we are concerned with subsets which contain boundaries which halve the amount of supersymmetry allowed. Specifically, we will work on the upper half plane $\Sigma=\mathbb{R}\times [0,\infty)$. At the boundary the left and right fermionic variables are related to each other. There are two ways to do this \cite{hori00}

\begin{equation}\label{boundarytype}
\begin{split}
&(A) \ \ \ \ \theta^+ +\operatorname{e}^{i\alpha}\bar\theta^-=0, \ \ \bar\theta^+ +\operatorname{e}^{-i\alpha}\theta^-=0,\\
&(B) \ \ \ \ \theta^+ -\operatorname{e}^{i\beta}\theta^-=0, \ \ \bar \theta^+ -\operatorname{e}^{-i\beta}\bar\theta^-=0.\\
\end{split}
\end{equation}

In theories with \emph{A-boundary} or \emph{B-boundary} the following supercharges are conserved, respectively:
\begin{equation}\label{abgenerators}
\begin{split}
&(A) \ \ \ \ \xbar Q_A:=\xbar Q_+ + \operatorname{e}^{i\alpha}Q_-, \ \ Q_A:=Q_+ +\operatorname{e}^{-i\alpha}\xbar Q_-,\\
&(B) \ \ \ \ \xbar Q_B:=\xbar Q_+ +\operatorname{e}^{i\beta}\xbar Q_-, \ \ Q_B:= Q_+ +\operatorname{e}^{-i\beta} Q_-.
\end{split}
\end{equation}

We say a theory has \emph{A-type supersymmetry} when it has conserved charges $(Q_A,\xbar Q_A)$ and A-boundary; and \emph{B-type supersymmetry} for  $(Q_B,\xbar Q_B)$ and B-boundary. In this paper we restrict to the case $\beta =0$ and $\alpha=0$ without loss of generality. The cases with general phases follow from the results below by using the $U(1)_V$ R-symmetry which maps the fermionic variables as $\theta^\pm\rightarrow e^{-i\alpha}\theta^\pm$, $\theta^\pm\rightarrow e^{-i\alpha}\theta^\pm$, and the superpotential as $W\rightarrow e^{-2i\alpha}W$.  In the present case, the B-type variation is given by
\begin{equation}\label{bvariation}
\delta_B=\epsilon Q_B- \bar \epsilon\  \xbar Q_B,
\end{equation}
which is a special case of the full $(2,2)$ variation where we take $\epsilon_+=-\epsilon_-=:\epsilon$ and $\bar \epsilon_+ = - \bar \epsilon_-=:\bar \epsilon$. The B-type generators obey the relations
$\left\{ Q_B,\xbar Q_B \right\} =-2i(\partial_+ +\partial_-)\ , \ Q_B^2=\xbar Q_B^2=0$. The general A-type variation \index{A-type supersymmetry} is given by
\begin{equation}\label{avariation}
\delta_A=\epsilon \bar Q_A -\bar \epsilon Q_A,
\end{equation}
Observe that the full $(2,2)$ variation preserves A-type boundary conditions if instead we take $\epsilon_+=\bar \epsilon_-=:\epsilon$ and $\bar \epsilon_+= \epsilon_-=:\bar \epsilon$.

The BRST operator $Q=Q_I$, where $I = A$ for A-type, and $I = B$ for B-type, is used to define the physical operators of the theory as those operators which are $Q$-closed.

\subsection{B-type defects}\label{btypeDefects}

In this section we specialize to B-type supersymmetry and review the use of matrix factorizations to describe B-supersymmetric boundary conditions and interfaces. Under the B-supersymmetry variation in equation (\ref{bvariation}) the action varies as $\delta_B S=\delta_B S_D+\delta_B S_F$. As noted in \cite{brunner03}, boundary terms may be added to cancel $\delta_B S_D$ at the boundary $\partial \Sigma$ (see also \cite{kapustin02, Orlov:2003yp} for a general description of topological B-type D-branes in LG models).  The variation of the $F$-term gives \cite{hori00}
\begin{equation}\label{varyf}
\delta_B S_{ F}=2i\epsilon\displaystyle\int_{\partial \Sigma}dtd\theta\  W( X)\big |_{\bar \theta=0}-2i\epsilon\displaystyle\int_{\partial \Sigma}dtd\bar \theta\ \xbar W( \xbar X)\big |_{ \theta=0},
\end{equation}
which is in general non-zero. To recover supersymmetry new boundary superfields are introduced on $\partial \Sigma$ \cite{govindarajan06}. These boundary superfields, which we denote by $\left\{\Pi_i\right\}_{i=1,\dots,r}$,  are fermionic and not chiral:
\begin{equation}\label{bdyfield}
\xbar D \Pi_i = E_i(X_{\partial}) \neq 0,
\end{equation}
where $\xbar D$ is the B-type covariant derivative and  $X_\partial$ denotes the boundary superfield associated to the bulk superfield $X$. The components of $\Pi$ are fermionic $\pi_i$ and scalar auxiliary fields $l_i$. The boundary superfields $\Pi_i$ carry the following action
\begin{equation}\label{bdyaction}
S_{\partial\Sigma}=\displaystyle \int dt  d^2 \theta \ \xbar \Pi_i \Pi_i + i\displaystyle \int_{\partial\Sigma} dt d\theta \ J_i\Pi_i\big |_{\bar \theta=0} - i\displaystyle \int_{\partial\Sigma}dt d\bar\theta \ \bar J_i\xbar\Pi_i\big |_{\theta=0},
\end{equation}
for some functions $J:=J\left(X_{\partial}\right)$. A more general form for the boundary coupling of B-type topological Landau-Ginzburg models is discussed in \cite{Lazaroiu:2003zi} but we do not use it here. The modified Landau-Ginzburg action in the topological twisted case is invariant under infinitesimal $B$-type supersymmetry variations iff $J_i E_i =W$ as functions of $X_\partial$.
The additional functional $S_{\partial\Sigma}$ provides a boundary contribution $Q_\partial$ to the BRST charge $Q$ that                                                                                             has the following form
\begin{equation}
Q_\partial =\sum_i J_i\pi_i +E_i\bar \pi_i .
\end{equation}

From the equation above, we see that different choices of the potentials $J_i$ and $E_i$ determine the Q-cohomology of the boundary fields. Choosing a representation for the Clifford algebra of the boundary fermions $\left\{\pi_i,\bar\pi_i\right\}_{i=1,\dots,r}$, we have \cite{brunner07}
\begin{equation}
Q_{\partial}= \begin{pmatrix} 0 & p_1 \\ p_0 & 0 \end{pmatrix},
\end{equation}
where the $p_i$ are $2^r \times 2^r$-matrices with polynomial entries in the chiral fields satisfying $p_1p_0 = p_0p_1 = W 1_{2^r \times 2^r}$ since $Q_\partial ^2 =W$. Thus the problem of characterizing the boundary spectra has been reduced to factorizing the superpotential over maps of arbitrary rank.

In general, given a polynomial $W\in S:=\mathbb{C}[X_j]$, and two $S$-modules $P_0$, $ P_1$, a matrix factorization of $W$ is an ordered pair $(p_0, p_1)$, where $p_i : P_i \rightarrow P_{i+1} \mod 2$, such that $p_i p_{i+1} = W 1_{i+1}$. One denotes matrix factorizations in the following way \cite{brunner07}
\begin{equation}\label{mfactor}
P=\left(  P_0 \mathrel{\mathop{\rightleftarrows}^{\mathrm{p_0}}_{\mathrm{p_1}}} P_1\right), \ \ \  p_0 p_1 =W 1_{P_1}, \ \ p_1 p_0 =W 1_{P_0}.
\end{equation}
The \emph{rank} of matrix factorization $P$ is the rank of the maps $p_i$. 

Aside from boundary conditions, matrix factorizations also describe defects. A defect is a one-dimensional interface that separates two Landau-Ginzburg models, or generally any two field theories. Here we take the defect to be at $x^1=y$, parallel to the $x^0$-axis. Let the region $x^1>y$ contain a LG model with superpotential $W_1(X_i)$, and the region $y>x^1>0$ contain a LG model with superpotential $W_2(Y_i)$.

Similar arguments as for the B-type boundary conditions shows that the degrees of freedom on the defect can be described by a matrix factorization of $W=W_1(X_i)-W_2(Y_i)$ over $\mathbb{C}[X_i,Y_i]$-modules. This is consistent with folding trick which says that such a defect is equivalent to the boundary condition of the tensored theory $\text{LG}_1\otimes\xbar{\text{LG}}_2$, where the bar means interchanging holomorphic and antiholomorphic variables \cite{brunner07}. The fusion of a defect and a boundary, or a defect and another defect is obtained by the tensor product of matrix factorizations. Let $P$ be the matrix factorization for a defect $D$ at $x^1=y$, and $Q$ for a boundary condition $B$ at $x^1=0$, given respectively by
\begin{equation}\label{defect}
P(X|Y)=\left( P_0 \mathrel{\mathop{\rightleftarrows}^{\mathrm{p_0}}_{\mathrm{p_1}}} P_1\right), \ \ \  p_0 p_1 =(W_1(X)-W_2(Y)) 1_{P_1}, \ \ p_1 p_0 =(W_1(X)-W_2(Y)) 1_{P_0},
\end{equation}
and
\begin{equation}\label{boundary}
Q(Y)=\left(  Q_0 \mathrel{\mathop{\rightleftarrows}^{\mathrm{q_0}}_{\mathrm{q_1}}} Q_1\right), \ \ \  q_0 q_1 =W_2(Y) 1_{Q_1}, \ \ q_1 q_0 =W_2(Y) 1_{Q_0}.
\end{equation}
Then a new boundary condition is obtained from their fusion, denoted
 \begin{equation}
B'=D*B,
\end{equation}
 by taking $y\rightarrow 0$. The resulting boundary is then given by the tensor product of matrix factorizations $Q'=P\otimes Q$ with
\begin{equation}\label{formaltensor}
Q'= \left( Q'_0=\left( P_0\otimes_{\mathbb{C}[Y]} Q_0\right)\oplus \left( P_1\otimes_{\mathbb{C}[Y]} Q_1\right) \mathrel{\mathop{\rightleftarrows}^{\mathrm{q'_0}}_{\mathrm{q'_1}}} \left( P_1\otimes_{\mathbb{C}[Y]} Q_0\right)\oplus \left( P_0\otimes_{\mathbb{C}[Y]} Q_1\right)=Q'_1\right),
\end{equation}
where,
\begin{equation}\label{mapfusion}
q'_0 = \begin{pmatrix} p_0\otimes 1_{Q_0} & 1_{P_1}\otimes q_1 \\ -1_{P_0}\otimes q_0 & p_1\otimes 1_{Q_1}  \end{pmatrix} \ \  , \ \
q'_1 = \begin{pmatrix} p_1\otimes 1_{Q_0} & -1_{P_0}\otimes q_1 \\ 1_{P_1}\otimes q_0 & p_0\otimes 1_{Q_1}  \end{pmatrix}.
\end{equation}

The tensor product above will give a result which is of infinite rank as a $\mathbb{C}[X]$-module. If the two initial defects are of finite rank, the infinity in the rank of the tensor product comes from trivial matrix factorizations which can be ``peeled off'' to obtain a reduced rank matrix factorization. To obtain the reduced rank matrix factorization resulting from equation (\ref{formaltensor}) more directly, one associates to each matrix factorization $P$ a 2-periodic $\mathbb{C}[X]/W$-resolution of the space $\coker p_1$, the cokernel of the $p_1$ map. Then the problem of computing $Q'$, the matrix factorization corresponding to the tensor product of $P$ and $Q$, is translated into finding $\coker q_1'$ in its reduced form. As noted in \cite{brunner07}, at the level of $\mathbb{C}[X]/W$-modules both $\coker q_1'$ and the space
\begin{equation}\label{trick}
V=\coker(p_1\otimes1_{Q_0}, 1_{P_0}\otimes q_1),
\end{equation}
have resolutions which are identical up to the last two steps. Therefore if we can find the reduced form of $V$, we can identify the 2-periodic resolution corresponding to the matrix factorization $Q'$. It turns out that it is simpler to work out the reduced form of $V$ since its components are the known maps of the original two matrix factorizations.


\subsection{A-branes and wave-front trajectories}

Similar to the previous section, we now specialize to A-type supersymmetry. The language of matrix factorizations lends itself naturally to be the B-type D-branes in Landau-Ginzburg models. This is not the case for D-branes preserving A-type supersymmetry. Below we give a geometric characterization of A-type D-branes, called A-branes for short.

We consider a general $\mathcal{N}=(2,2)$-supersymmetric sigma model in two dimensions  with superpotential $W$ defined on $\Sigma =\mathbb{R}\times[0,\infty)$, and with an $n$-dimensional target space $M$ which we assume to be a K\"ahler manifold. Let $\gamma \subset M$  contain the embedding of $\partial \Sigma$, that is $\phi:\partial\Sigma \hookrightarrow \gamma$ where $\phi$ denotes the lowest components of the superfields. Then, a D-brane wrapped on $\gamma$ preserves A-supersymmetry iff $\gamma$ is Lagrangian submanifold of $M$ with respect to the K\"ahler form, and $W(\gamma)\subset \mathbb{C}$ is a straight line parallel to the real axis, and invariant under the gradient flow of $\real W$ \cite{hori00}. A submanifold $N$ of a symplectic manifold $(M,\omega)$ is called \emph{Lagrangian} if the symplectic form $\omega$ vanishes on $N$, and $\dim N = 1/2 \dim M$.

An example of A-branes are those D-branes wrapped on the submanifold defined by the action of the gradient of $\real W$ on a nondegenerate critical point of the superpotential $W$. This case is discussed in \cite{hori00} and we will review it below, but first we introduce some terminology from complex variables. A non-constant holomorphic function $f$ has a \emph{critical point} at $z_0$ if $f'(z_0)=0$. The \emph{order} of the critical point is the order of zero of $f'$ at $z_0$. The value of $f(z_0)$ is called the \emph{critical value}.

 For definiteness, let $X_*$ be a critical point of $W$ of order $n=1$, and let $f_X(t)=f(t,X)$ be the global flow \index{global flow} generated by $\operatorname{grad}[\real W]$. In general a global flow is a continuous map $f:[0,1)\times M \rightarrow M$ which satisfies $f(0,X)=X$, $f(t,f(s,X))=f(t+s,X)$. Here we are interested in a flow that satisfies
\begin{equation}\label{flowdef}
f'_X(t)=\operatorname{grad}[\real W]_{f_X(t)},
\end{equation}
where $\operatorname{grad}[\real W] = g^{IJ}\partial_J (\real W)\partial_J$, and we evaluate this vector field at the point $f_X(t)\in M$.

Defining the``wave-front trajectory'' submanifold
\begin{equation}\label{gammadef}
\gamma_{X_*} :=\left\{ X\in M \big | \lim_{t\rightarrow -\infty} f_X(t)=X_*\right\},
\end{equation}
then the claim is D-branes wrapped on $\gamma_{X_*}$ are A-branes. This means checking that $\gamma_{X_*} $ is Lagrangian submanifold whose image in the $W$-plane is parallel to the real axis.
\begin{equation}
\begin{split}
\operatorname{grad}[\real W](\image W) &=g^{IJ}\partial_J(W+\xbar W)\partial_I (W-\xbar W)\\
&=g^{i\bar j}\partial_{\bar j}\xbar W\partial_i W - g^{\bar j i}\partial_iW \partial_{\bar j}\xbar W\\
&=|\partial W|^2-|\partial W|^2\\
&=0.
\end{split}
\end{equation}
Therefore $\image W$ is constant along $\operatorname{grad}[\real W]$ and thus $W(\gamma_{X_*})$ is a ray starting at the critical value $w_*:= W(X_*)$ and parallel to the real axis. The name ``wave-front trajectory'' for $\gamma_{X_*}$ follows from the constant value of $\image W$.

Now we need to show that $\gamma_{X_*}$ is middle dimensional. Recall that if $z_0$ is a critical point of $f:\mathbb{C}\rightarrow\mathbb{C}$ of order $m-1$, then there exists a change of coordinates near $z_0$ and $f(z_0)$ such that $f$ has the form $f(\xi)=\xi^m+f(z_0)$. There is an analogous statement for several complex variables (Complex Morse Lemma) that allows us to map a neighborhood of $X_*$ to that of $0\in\mathbb{C}^n$,
\begin{equation}\label{wlocal}
W=w_*+\sum_{i=1}^n z_i^2+ o(z_i^3).
\end{equation}

Suppose that this change of variables leaves us with a flat metric, $ds^2=\sum_i|dz_i|^2$.  Then if we write $z_i(t)$ for the components of the map $f_X(t)$, the flow equation (\ref{flowdef}), in a region sufficiently close to $0\in \mathbb{C}^n$ (so that we can ignore the higher order terms in $W$), becomes
\begin{equation}
\label{eq:FlatComponentFlow}
z'_i(t)=\bar{z}_i(t),
\end{equation}
or, breaking into real and imaginary parts, $z_i=x_i+iy_i$,
\begin{equation}
\label{eq:FlowSolutions}
x'_i(t)=x_i(t),\quad y'_i(t)=-y_i(t),\qquad\Rightarrow\qquad x_i(t)=X_ie^t,\quad y_i(t)=Y_ie^{-t},
\end{equation}
where $X_i$ and $Y_i$ are simply the coordinates of the point $X$.  The submanifold $\gamma_{0}$ is then determined by those points $X$ which satisfy $f_X(t)\rightarrow 0$ as $t\rightarrow -\infty$.  Looking at our solutions (\ref{eq:FlowSolutions}), these are simply the points with $Y_i=0$, $X_i$ arbitrary.  In particular, near $0$ (i.e.\ in a small neighborhood of $X_\ast$), $\gamma_{X_\ast}$ is an $n$-dimensional real submanifold.

Generally we won't be lucky enough that the metric is flat.  However, restricting to a sufficiently small neighborhood of $0$, we can assume that the metric is constant and hermitian.  Since any positive-definite hermitian matrix can be connected to the identity matrix by a continuous path in the space of positive-definite hermitian matrices, one can show that the matrix $M_a^{\hphantom{a}b}$ appearing in the flow equation (now written in real components)
\begin{equation}
x'_a(t)=M_a^{\hphantom{a}b}x_b(t),
\end{equation}
will always have $n$ positive and $n$ negative eigenvalues, and hence $\gamma_0$ remains middle-dimensional.

We are left to show that the induced symplectic form vanishes on $\gamma_{X_*}$. Defining $v:= \operatorname{grad}[\real W]$ one can compute the Lie derivative result $\mathcal{L}_v \omega=0$ (true whenever $v$ is the gradient of a holomorphic plus an antiholomorhic function), which means that $\omega$ is invariant along the gradient of $\real W$.   Now we can use this fact to show that for all $X\in\gamma_{X_*}$ $\omega_X (v_1,v_2)=0 $;  where $v_1, v_2 \in  T_X\gamma_{X_*}$, the tangent space to $\gamma_{X_\ast}$ at $X$. Indeed, since $\omega$ is invariant along the flow $f(t,X)=f_t(X)$ generated by the vector field $v$, it follows that $(f^*_t\omega)_X:=f^*_t(\omega_{f(t,X)})$ equals $\omega_X$ for all $X$ and $t$. Therefore, $\omega_X (v_1,v_2)= f^*_t( \omega_{f_X(t)})(v_1,v_2)= \omega_{f_X(t)}(f_{t*} v_1,f_{t*}v_2)$. For any function $g$ on $M$ and any $X\in\gamma_{X_*}$,
\begin{equation}
\lim_{t\rightarrow -\infty} (g \circ f_t)(X)=g(X_*),
\end{equation}
so $g\circ f_t$ becomes a constant function along $\gamma_{X_*}$. It then follows that $f_{t*}v\rightarrow 0$ for $v\in T_X\gamma_{X*}$.

Hence $\omega_X (v_1,v_2)=0 $ since it is independent of the parameter $t$. Thus $\gamma_{X_*}$ is a Langrangian submanifold of $(M,\omega)$. To summarize, we have shown that D-branes wrapped on $\gamma_{X_*}$ as defined in equation (\ref{gammadef}) are A-branes which are mapped to $W(X_*)+\mathbb{R}^{\geq 0}$, where $X_*$ is a critical point of $W$.


\subsection{A-branes in Landau-Ginzburg models}\label{abraneslg}

 We use the wave-front trajectory example for LG models with polynomial superpotentials. We first consider the case $W=X^{k+2}$ with $k$ a non-negative integer.  Then $W$ has only one critical point $X_*=0$. As noted above we know that $\gamma_{0}$ (defined in (\ref{gammadef})) is the preimage of the set $[0,\infty)\subset\mathbb{C}$. Explicitly, A-branes wrap the submanifold
\begin{equation}
\gamma_{0}=\left\{ r \exp\left(  \frac{2\pi ni}{k+2}\right) : r\in[0,\infty) \ , \ n \in \left\{ 0,\dots,k+1\right\} \right\}\subset\mathbb{C}.
\end{equation}

Using submanifolds of $\mathbb{C}$ which asymptote to $\gamma_0$, we can also describe the A-branes of LG theories with more general superpotentials of the type
\begin{equation}\label{deformedlg}
W_{ \lambda}(X)=X^{k+2}+\sum_{j=0}^{k-1} \lambda_j X^{j+2}.
\end{equation}
We have observed that a constant term does not contribute to the fermionic integral of the Lagrangian so it can be shifted away. A linear term does not introduce any new branch points. So we have the freedom to gauge it away and thus always translating one of the critical points to the origin.

In the most general case, $\lambda_j\neq0$ for all $j$, and $W_{ \lambda}$ has $k+1$ non-degenerate critical points which are isolated. In this case we have $k+1$ possible Lagrangian submanifolds to wrap the A-branes, corresponding to each of the critical points. We assume that $\image w_i\neq \image w_j$ for $i\neq j$, where $w_j:=W_{ \lambda}(X_{*j})$ are the critical values. This assumption eliminates the possibility of having overlapping images in the $W$-plane of the submanifolds $\gamma_{i}$ corresponding to the $X_{*j}$ critical points.

The A-branes of the deformed theory are curves asymptoting to $\mathcal{L}_{n_1}\cup \mathcal{L}_{n_2}$, $n_1\neq n_2$, where $\mathcal{L}_{n_j} \subset \gamma_{0}$ are slices corresponding to each value of $n_i \in\left\{ 0,\dots,k+1\right\}$. This claim follows by noting that for large $X$, $W_\lambda$ approaches the undeformed $W$ since the leading term $X^{k+2}$ in $W_\lambda$ dominates. So $W^{-1}_\lambda$ is close to $W^{-1}$ in this regime. Now, let $X_{*j}$ be one of the critical points of the deformed potential. By assumption it is of order one so locally near $X_{*j}$ and its image, $W_{\lambda}$ is biholomorphically equivalent to a quadratic map. Thus the preimage of $w_j+\mathbb{R}^{\geq 0}$ near $w_j$ is two wavefront trajectories starting at $X_{*j}$. As noted, these curves approach some $\mathcal{L}_{n_1}$ and $\mathcal{L}_{n_2}$. The curves intersect at the branch points only (consider $W_\lambda$ as a branched cover) which means $n_1\neq n_2$. For non-generic values of the $\lambda_j$, the branch points can be degenerate. Then the A-brane associated with one of these points, say $X_*$, will asymptote  $\mathcal{L}_{n_1}\cup\cdots \cup \mathcal{L}_{n_{o(X_*)+1}}$, where $o(X_*)$ is the order the critical point $X_*$.

Following the work of \cite{brunner07a} we can depict the A-brane description above for the Landau-Ginzburg models by compactifying the $X$-plane to the disk $D$. The resulting graph contains the critical points $X_{*i}$ in the interior of the disk; cyclically ordered preimages $\left\{B^1,\dots, B^{k+2}\right\}$ of $\infty\in W$-plane on the boundary of the disk $\partial D$; and $({o(X_{*i})+1})$-many segments $\gamma_i^a$ connecting the point $X_{*i}$ to that many of the $B^a$. We define $\Gamma_i := \cup_a  \gamma_i^a$ and $\Gamma := \cup_i \Gamma_i$. We call the graph formed by $\Gamma$ and the boundary $\partial D$ the \emph{schematic representation} of the superpotential. The two graphs below are examples of schematic representations for A-branes in LG models with superpotentials $W=X^4$ and $W=X^4+\lambda X^3$.

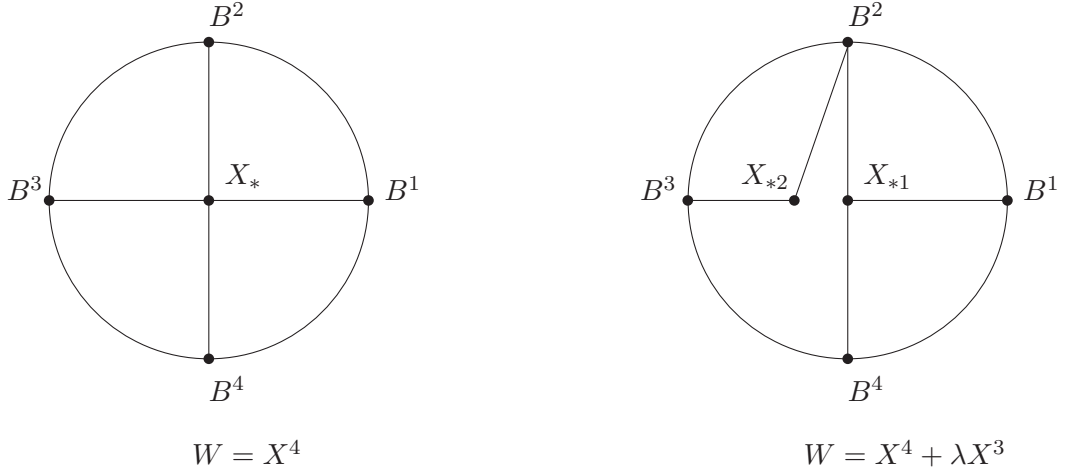
\setlength{\unitlength}{7mm}
\begin{picture}(10,10)(-6,-6)

\put(0,0){\circle{6}}
\put(0,0){\line(0,5){3}}
\put(0,0){\line(0,-5){3}}
\put(0,0){\line(5,0){3}}
\put(0,0){\line(-5,0){3}}
\put(0,0){\circle*{.2}}
\put(3,0){\circle*{.2}}
\put(-3,0){\circle*{.2}}
\put(0,3){\circle*{.2}}
\put(0,-3){\circle*{.2}}
\put(.3,.3){$X_*$}
\put(3.3,0){$B^1$}
\put(0,3.3){$B^2$}
\put(-3.8,0){$B^3$}
\put(0,-3.8){$B^4$}
\put(-.5,-5){{ $W=X^4$}}

\put(12,0){\circle{6}}
\put(12,0){\line(0,5){3}}
\put(12,0){\line(0,-5){3}}
\put(12,0){\line(5,0){3}}
\put(11,0){\line(-5,0){2}}
\put(11,0){\line(1,2.9){1}}
\put(10,.3){$X_{*2}$}
\put(11,0){\circle*{.2}}
\put(12,0){\circle*{.2}}
\put(15,0){\circle*{.2}}
\put(9,0){\circle*{.2}}
\put(12,3){\circle*{.2}}
\put(12,-3){\circle*{.2}}
\put(12.3,.3){$X_{*1}$}
\put(15.3,0){$B^1$}
\put(12,3.3){$B^2$}
\put(8.1,0){$B^3$}
\put(12,-3.8){$B^4$}
\put(11,-5){{ $W=X^4+\lambda X^3$}}

\end{picture}

A graphical representation $\Gamma$ has the following properties \cite{brunner07a}: all the preimages of a critical value $\omega\in \mathbb{C}$ are connected on $\Gamma$;   $\Gamma \setminus\partial D$ is connected and simply connected;  $\forall i\neq j, \Gamma_i \cap \Gamma_j$ contains at most one point; and it is non-empty only if it contains an element of the fiber $f^{-1}(\infty)$;  $\Gamma_i \cap \Gamma_j\cap \Gamma_k = \emptyset$.

\section{Describing RG flows in $\mathbb{C}/\mathbb{Z}_d$ orbifolds using defects}

In this section we describe a new way of dealing with the $\mathbb{C}/\mathbb{Z}_d$ orbifold in terms of defects. The language of matrix factorizations can be utilized to describe the RG flow between the $\mathbb{C}/\mathbb{Z}_d$ orbifolds. This can be done directly by considering the Lagrangian of the model as equivalent to that of a LG model with superpotential $W=0$. So any defects between $\mathbb{C}/\mathbb{Z}_m$ and $\mathbb{C}/\mathbb{Z}_n$ become a problem of factorizing the zero polynomial.

Since we are working with B-type supersymmetry we need to use a perturbation which preserves this type. Such a perturbation for a $\mathcal{N}=(2,2)$ theory is done using twisted chiral fields $\Psi$ in theory with the integrals
\begin{equation}
\Delta S = \displaystyle \int_\Sigma d^2x d\bar x^- d\theta^+ \ \Psi \big|_{\bar \theta^+ = \theta^- =0}.
\end{equation}
But the $\mathcal{N}=(2,2)$ supersymmetry dictates that the parameters of the superpotential and twisted superpotential remain decoupled under the RG flow \cite{hori02}. This fact means that the structure of the twisted chiral sectors is independent of the specific superpotential. Especially in our case whether there is one or not. Therefore the spectrum of the twisted chiral sectors between  $\mathbb{C}/\mathbb{Z}_d$  and the $\mathbb{Z}_d$-orbifolded LG with $W=X^d$ are equivalent, and their B-type preserving perturbations can be mapped to each other.  With this observation we set out to check that the sort of defects presented in \cite{brunner07a} describing the RG flow defects coming from such perturbations over a subset of $\Sigma$, can be extended to the non-compact orbifolds and the RG flows between them.


\subsection{$\mathbb{C}/\mathbb{Z}_d$ as an $\text{LG}/ \mathbb{Z}_d$ with $W=0$}\label{rgflowDefects}

Superstring theory on the space $\mathbb{C}/\mathbb{Z}_d$ can be described by a chiral superfield
\begin{equation}
\Phi=\phi(y^{\pm})+\theta^\alpha\psi_\alpha(y^\pm)+ \theta^+\theta^- F(y^\pm),
\end{equation}
where $y^\pm=x^\pm-i\theta^\pm\bar\theta^\mp$.  The action takes the form
\begin{equation}\label{LGnoW}
S=\displaystyle  \int d^2 x d^4\theta \ \xbar \Phi \Phi +0,
\end{equation}
where we included the zero to emphasize that we have a LG model with superpotential $W=0$ in the D-term. In this way we can construct defects between different  $\mathbb{C}/\mathbb{Z}_d$ orbifolds and describe them in terms of matrix factorizations. Indeed, we check that when two $\mathbb{C}/\mathbb{Z}_d$  theories are related by an RG flow, we can juxtapose them with a corresponding defect which maps the boundary conditions accordingly.

Matrix factorizations of the zero polynomial work in exactly the same way as the case for any other polynomial. As an example of this we consider the fusion of a defect between two orbifolded theories; the upper one with superpotential $W_1(X)=X^d$ and the lower one with $W_2(Y)$ the zero superpotential but orbifold group $\mathbb{Z}_{d'}$. The simplest such defect is given by

\begin{equation}\label{defectNoW}
D^{m,n,N}(X|Y)=\left(  D_1=  \mathbb{C}[X,Y][m,-n] \mathrel{\mathop{\rightleftarrows}^{ X^N}_{X^{d-N}}} \mathbb{C}[X,Y][m-N,-n]=D_0\right),
\end{equation}
where $[\cdot, \cdot]$ is the $\mathbb{Z}_d \times\mathbb{Z}_{d'}$ grading. We see that $d_1 d_0 = X^d-0 = W_1(X)-W_2(Y)$. In the lower theory, the boundary conditions corresponding to rank-1 matrix factorizations are a direct sum of the irreducible matrix factorizations of the form
\begin{equation}\label{boundaryNOw}
Q^{L,M}(Y)=\left(  Q_1=  \mathbb{C}[Y][L+M] \mathrel{\mathop{\rightleftarrows}^{ Y^M}_{0}} \mathbb{C}[Y][L]=Q_0\right),
\end{equation}
where $L\in\mathbb{Z}_{d}$ labels the irreducible representations.

If the defect $D^{m,n,N}$ sits at $x^1=y$ and we take $y\rightarrow 0$ the fusion of the defect and the boundary condition is given by tensor product of both matrix factorizations. This is obtained by looking at $\coker f = D_0\otimes Q_0/ \operatorname{im} f$ where $f=(d_1\otimes1_{Q_0}, 1_{D_0} \otimes q_1)$ \cite{enger05}. We denote the $\mathbb{C}[X,Y]$-generators of $D_0$ and $Q_0$ by $e^{D_0}_{m,n}$ and $e^{Q_0}_{L}$, respectively. Then as a $\mathbb{C}[X]$-module, $\coker f$ is generated over $e^i:=Y^i e^{D_0}_{m,n}\otimes e ^{Q_0}_{L}$ modulo
\begin{equation}
X^N e^i=0\ \ \ , \ \ \ e^{i+M}= 0, \ \ \ \forall i\geq 0.
\end{equation}
The second condition means that $V$ has rank $M$. Note that $e^i$ has $\mathbb{Z}_d \times\mathbb{Z}_{d'}$-degree $[m-N,L-n+i]$, but under fusion we are left with a $\mathbb{Z}_{d}$ theory so we have to extract the $\mathbb{Z}_{d'}$-invariant subset $V^{\mathbb{Z}_{d'}}\subset V$. This means the $i$ is fixed to $i=n-L$, which means we are left with one generator with $\mathbb{Z}_{d}$-degree $m-N$ restricted to $X^N=0$.  Otherwise if $n-L \not\in [0,M-1]$ then $D^{m,n,N}*_{\text{orb}}Q^{L,M}=0$. In summary,

\begin{equation}\label{example1}
D^{m,n,N}*_{\text{orb}}Q^{L,M}= \left\{ \begin{array}{rl}
 Q^{m,N}, &\mbox{ if $n-L\leq M-1$,} \\
  0, &\mbox{ otherwise.}
       \end{array} \right.
\end{equation}

Another example of useful defects given by matrix factorizations of $W=0$ are those enforcing the action of the symmetry group. Similar to those in \cite{brunner07a} they are given by the $\mathbb{Z}_d\times\mathbb{Z}_d$- equivariant matrix factorization $T^m = (T^m_1, T^m_0; t_1, 0)$ with
\begin{equation}
T^m_1 = {}_{\mathbb{C}[X,Y]}\left\{ e^{1}_{m,k}\right\}_{(m,k) \in \mathbb{Z}_d\times\mathbb{Z}_d} \ \ , \ \ \deg e^1_{m,k} =[m+k+1,-k],
\end{equation}
\begin{equation}
T^m_0 = {}_{\mathbb{C}[X,Y]}\left\{ e^{0}_{m,k}\right\}_{(m,k) \in \mathbb{Z}_d\times\mathbb{Z}_d} \ \ , \ \ \deg e^0_{m,k} =[m+k,-k].
\end{equation}
The factorizing map is given by
\begin{equation}
t_1=\sum_{k=0}^{d-1} \left( X e_{m,k}^0\otimes {e^1}_{m,k} ^*- Y e_{m,k+1}^0\otimes {e^1}_{m,k} ^*\right),
\end{equation}
where $e^*$ is the basis dual to $e$.

One obtains the fusion rules
\begin{equation}
T^m *_{\text{orb}} T^n=T^{m+n},
\end{equation}
and
\begin{equation}
T^m *_{\text{orb}} Q^{M,N}=Q^{M+n ,N},
\end{equation}
where $D_1*_{\text{orb}} D_2$ means extracting the part of $D_1 * D_2$ which is invariant under the symmetry group of the theory between both defects $D_1$ and $D_2$. The sums are performed modulo $d$. Hence the defects $T^m$ form a representation of the symmetry group.

More importantly, we note that by also setting $p_0=0$ in the special defects introduced in \cite{brunner07a} we obtain defects which act as the interface between orbifolds sitting at opposite endpoints of the RG flow. The special defects are $\mathbb{Z}_{d'}\times\mathbb{Z}_d$ - equivariant matrix factorizations $P^{(m,\underline n)}$, with labels $m\in \mathbb{Z}_d$ and $n=(n_0,\dots, n_{{d'}-1})$ with $n_i\in\mathbb{N}_0$ such that $\sum_i n_i=d$. The $\mathbb{C}[X,Y]$-modules $P_1$ and $P_0$ and their  $\mathbb{Z}_{d'}\times\mathbb{Z}_d$-grading are given by,
\begin{equation}
P_1 = \mathbb{C}[X,Y]^{d'}
\begin{pmatrix} [1,-m] \\ [2,-m-n_1]  \\ [3,-m-n_1-n_2 ]\\ \vdots \\ [d', -m-\sum_{i=1} ^{d'-1}n_i ]\end{pmatrix} \ \ \ , \ \ \ P_0 = \mathbb{C}[X,Y]^{d'}
\begin{pmatrix} [0,-m] \\ [1,-m-n_1]  \\ [2,-m-n_1-n_2 ]\\ \vdots \\ [d'-1, -m-\sum_{i=1} ^{d'-1}n_i ]\end{pmatrix}.
\end{equation}
The factorizing maps are
\begin{equation}\label{mapwithzero}
p_1^{m,n}=Y 1_{d'}- \Xi_n(X)\ \ \ , \ \ \ p_0^{m,n} = 0 ,
\end{equation}
where $(\Xi_n(X))_{a,b}:=\delta^{(d')}_{a,b+1}X^{n_a}$.

As computed in \cite{brunner07a} the general rule for fusion of a special defect $P^{(m,n)}$ and a $\mathbb{Z}_d$-irreducible boundary condition $Q^{(M,N)}$ is
\begin{equation}\label{genfusion}
P^{(m,\underline n)}*Q^{(M,N)}=\bigoplus_{a\mathbb{Z}_{d'}:\ i(a)<\text{min}(N,n_a)} Q^{(a,k(a))},
\end{equation}
where $i(a)=\left\{n-M+\sum_{j=0}^an_j\right\}_d$.

One can check that special defects send the boundary condition $Q^{(M,1)}$ to another such boundary condition with $N=1$, $Q^{(M',1)}$.

Let $P^{(m,\underline n)}$ be a special defect and $Q^{(M,N=1)}$ an irreducible B-type boundary condition. Then their fusion is

\begin{equation}\label{defectbdy}
P^{(m,\underline n)}*Q^{(M,N=1)}=\begin{cases} 0, &  M\notin \mathcal{L}_{(m,\underline n)}\\ Q^{(a,1)}, & M=m+\sum_{i=1}^a n_i
\end{cases}
\end{equation}
where $\mathcal{L}_{(m,\underline n)}:=m+\left\{ n_0, n_0+n_1,\dots,n_0+n_1+\cdots n_{d'-1}\right\}$.

\subsection{Comparison with RG flow in the $\mathbb{C}/\mathbb{Z}_d$ theories}
We can compare the result for the fusion of the defects $P^{m,n}$ with boundary conditions $Q^{M,N}$ of the LG model with zero superpotential with the RG flow between the $\mathbb{C}/\mathbb{Z}_d$ orbifolds. 
 For this purpose we describe the RG flow in these models by looking at their chiral rings.

Upon bosonizing the fermionic fields of the superstring theory, one can construct the chiral operators given in \cite{harvey01}
\begin{equation}
X_j = \sigma_{j/n}\exp[i(j/n)(H-\xbar H)] \ \ , \ \ j =1,\dots, n-1,
\end{equation}
where $\sigma_{j/n}$ is the bosonic twist operator. These operators are the bosonic components of the respective chiral fields which we will also denote by $X_j$. The higher chiral fields are powers of $X:=X_1$. The chiral ring of this theory is generated by $X$ and
\begin{equation}
Y:=\frac{1}{V_2}\psi\psi= \frac{1}{V_2}\exp[i(H-\xbar H)],
\end{equation}
modulo
\begin{equation}\label{ringrel}
X^d=Y.
\end{equation}

Deformations of equation (\ref{LGnoW}) by the following F-term preserve supersymmetry since the $X_j$ fields are chiral,
\begin{equation}\label{deflangrangian}
\delta L = \sum_{j=1} ^{n-1} \lambda^j \displaystyle \int  d^2\theta \ X_j.
\end{equation}
The deformed theory has a chiral ring with the same fields as before but with relation in equation \ref{ringrel} altered to
\begin{equation}\label{deformed}
X^d+\sum_{j=1} ^{d-1}g_j(\lambda)X^j=Y,
\end{equation}
where $g_j(\lambda)$ are polynomials in the couplings \cite{harvey01}. A deformation such as in equation (\ref{deflangrangian}) induces a RG flow in the theory. By considering the case where $g_i=0$ for $i\leq d'-1$, the IR and UV limits of the ring condition above are $X^d=Y$ and $g_{d'}X^{d'}=Y$ respectively. These two are the conditions defining $\mathbb{C}/\mathbb{Z}_d$ and $\mathbb{C}/\mathbb{Z}_{d'}$, respectively.

We note that for every RG flow $\mathbb{C}/\mathbb{Z}_d \longrightarrow \mathbb{C}/\mathbb{Z}_{d'}$ there exists a matrix factorization $P^{(m,\underline n)}$ of $W=0$ representing a defect $D$ between $\mathbb{C}/\mathbb{Z}_d$ and $\mathbb{C}/\mathbb{Z}_{d'}$. Given two such bulk theories, we can juxtapose them via a defect $P^{(m,\underline n)}$ by choosing $m\in\mathbb{Z}_d$ and non-negative integers $\left\{n_0,n_1,\dots,n_{d'-1}\right\}$ subject to $n_0+\cdots n_{d'-1}=d$. The solution is a non-unique defect but that reflects the action of the overall $\mathbb{Z}_{d'}\times \mathbb{Z}_{d}$ symmetry. In the next section we will have a better description of how the boundary degrees of freedom are mapped from one theory to the other under fusion with RG flow defects.

As an example, consider the $\mathbb{Z}_5$ orbifold. In this case the chiral ring of the deformed theory is defined modulo $X^5+\sum_{j=1}^4 g_j(\lambda)X^j=Y$. If we set $g_1=g_2=0$, then the RG flow goes between $\mathbb{C}/\mathbb{Z}_5$ in the UV limit (since the theory's chiral ring has the relation $X^5=Y$) and  $\mathbb{C}/\mathbb{Z}_3$ (since in the IR limit the defining relation is $X^3=Y$). Then the defect $P^{(3,\underline n)}$ with $\underline n = (2,2,1)$ can sit at the interface  between the theories $\mathbb{C}/\mathbb{Z}_5$ and $\mathbb{C}/\mathbb{Z}_3$ such that B-type supersymmetry is preserved across the interface.

\section{RG flows using mirror models}

A second strategy is to study the orbifold RG flow
in terms of the mirror of  $\mathbb{C}/\mathbb{Z}_d$ \cite{vafa01}. Using mirror symmetry we obtain the diagram below. In the following $m$ stands for mirror symmetry and $|_B$  for the B-type defects; $\text{LG}_m$ denotes the LG model with $W=X^m$;  and $\widetilde{\text{LG}}_m$ the twisted LG with $W=\widetilde X^m$. 

\begin{equation}\label{modeldiagram}
\begin{CD}
@. \text{LG}_m/\mathbb{Z}_m @ > \big |_B>> \text{LG}_n/\mathbb{Z}_n\\
@. @VV\cong V@ VV\cong V  @.\\
\mathbb{C}/\mathbb{Z}_m @ >m>> \widetilde{\text{LG}}_m @ . \ \ \ \widetilde{\text{LG}}_n @>m>> \mathbb{C}/\mathbb{Z}_n @ . \\
@. @VV m V@ VV m V  @.\\
@. \text{LG}_m @> RG >> \text{LG}_n\\
\end{CD}
\end{equation}

In the diagram above, the mirror mapping from   $\mathbb{C}/\mathbb{Z}_n$ to a twisted LG with non-vanishing potential comes from a mirror correspondence between a gauged linear sigma model (GLSM) and a more general LG theory. As detailed in \cite{vafa01,Hori:2000kt}, one considers a GLSM whose geometry is described by 
\begin{equation}
-d|X_0|^2 + \sum_{i=1}^n k_i |X_i|^2=t,
\end{equation}
where the fields $(X_0, X_i)$ come with $U(1)$ charges $(-d,k_i)$, and $t$ is the complexified Fayet-Iliopoulos (FI) parameter. Such GLSM is mirror to a LG theory with superpotential
\begin{equation}
\widetilde W=\sum_{i=1}^n Z_i^d +e^{t/d}\prod_{j=1}^n Z_j^{k_j},
\end{equation}
where the variables $Z_i$ are twisted chiral fields, and the superpotential is taken modulo $(\mathbb{Z}_d)^{n-1}$. The IR fixed point of the GLSM is obtained with the limit $t\rightarrow - \infty$. This limit breaks the $U(1)$ symmetry to $\mathbb{Z}_d$ and the geometry obtained is that of $\mathbb{C}^n/\mathbb{Z}_d$. In this note we consider the $n=1$ case, i.e. $\mathbb{C}/\mathbb{Z}_d$. On the mirror side, the $t\rightarrow - \infty$ limit gives us the LG with $\widetilde W= Z^d$. Thus we see that the RG flow between the non-compact orbifolds can be described in terms of matrix factorizations of true LG orbifolds with non-zero superpotentials.

\subsection{RG flow defects using mirror models}

The idea is that via mirror symmetry we can represent the $\mathbb{C}/\mathbb{Z}_d$ orbifold as a twisted LG model with superpotential $W=\widetilde X^d$. We denote this theory  by $\widetilde{ \text{LG}}_d$ in the above diagram. This theory is equivalent to the model $\text{LG}_d/\mathbb{Z}_d$,  the orbifold of a non-twisted LG model with superpotential $W=X^d$ by $\mathbb{Z}_d$. So we can use defects between these LG orbifolds to study the RG flow between the original $\mathbb{C}/\mathbb{Z}_d$ orbifolds.

As in the previous section we are again in the Landau-Ginzburg model so we can use the RG flows defects $P^{(m,\underline n)}$. The factorizing maps are as in equation (\ref{mapwithzero}) but with $p_0$ non-zero:

\begin{equation}
p_1^{m,n}=Y 1_{d'}- \Xi_n(X)\ \ \ , \ \ \ p_0^{m,n} = \prod_{i=1}^{d'-1}(Y 1_{d'}- \eta^i\Xi_n(X)),
\end{equation}
where $\eta$ is an elementary $d'th$ root of unity. And similarly, the irreducible matrix factorizations corresponding to these boundary conditions are of the same form as in equation (\ref{boundaryNOw}),
\begin{equation}\label{}
Q^{L,M}(Y)=\left(  Q_1=  \mathbb{C}[Y][L+M] \mathrel{\mathop{\rightleftarrows}^{ X^M}_{X^{d'-M}}} \mathbb{C}[Y][L]=Q_0\right).
\end{equation}

We review the graphical version introduced in  \cite{brunner07a} to depict the fusion of $P^{(m,\underline n)}$ with the boundary conditions $Q^{(L,M)}$. To the set $Q^{(\underline M,1)}:=\left\{Q^{(M,1)} :0\leq M\leq d-1\right\}$ the following graph is assigned: A disk divided into $d$ equal sections by segments from the origin to the boundary. One segment is decorated to start labeling the sections $S_i$ from $i=0$ to $s=d-1$. Below is such a graph for $d=4$:

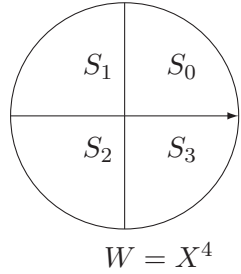
\setlength{\unitlength}{5mm}
\begin{center}
\begin{picture}(10,10)(-6,-6)
\put(0,0){\circle{6}}
\put(0,0){\line(0,5){3}}
\put(0,0){\line(0,-5){3}}
\put(0,0){\vector(5,0){3}}
\put(0,0){\line(-5,0){3}}
\put(1.1,1.1){$S_0$}
\put(-1.1,1.1){$S_1$}
\put(-1.1,-1.1){$S_2$}
\put(1.1,-1.1){$S_3$}
\put(-.8,-4){{ $W=X^4 $}}
\end{picture}
\end{center}


Using the graphical description described above, the special defects $P^{(m,n)}$ are represented by the operators
\begin{equation}\label{graphspecial}
\mathcal{O}^{(m,n)}:=\mathcal{T}_{-a(m,n)}\mathcal{S}_{\mathcal{L}^c(m,n)},
\end{equation}
where $\mathcal{L}_{(m,n)}$ is defined below (\ref{defectbdy}) and $a(m,n):=|\left\{0,\dots,m\right\}\cap \mathcal{L}_{(m,n)}|$. The operator $\mathcal{S}_{\left\{s_1,\dots, s_k\right\}}$  deletes the sectors $S_{s_j}$ by merging the segments which bound them. The operator $\mathcal{T}_k$ acts as the $\mathbb{Z}_d$-symmetry by shifting $M\rightarrow M+k$ in $Q^{(M,1)}$. So just like $P^{(m,\underline n)}$, the operator $\mathcal{O}^{(m,n)}$ annihilates the sectors associated to boundary conditions whose label $M$ does not belong in $\mathcal{L}_{(m,n)}$. Then it relabels the remaining sectors by setting the $S_{m}$ to $S_0$.

The above pictorial representation generalizes to boundary conditions $Q^{(M,N)}$ with $N>1$ as well. In this case, $Q^{(M,N)}$ corresponds to the union $S_M\cup S_{M+1}\cup\dots\cup S_{M+N-1}$. We want to show that the operators in the definition (\ref{graphspecial}) still represent the action of special defects on the boundary conditions in this $N>1$ case.

Represent $Q^{(M,N)}$ by $S^{(M,N)}:=S_{M}\cup\dots\cup S_{M+N-1}$ and assume that $\mathcal{S}_{\mathcal{L}^c_{(m,n)}}$ shrinks $S^{(M,N)}$ to nothing. Then $\left\{M,M+1,\dots,M+N-1\right\}\subset \mathcal{L}^c_{(m,n)}$. Thus, $M+k\neq m +\sum_{i=1}^a n_i \ \forall a \in \mathbb{Z}_{d'},\ N-1\geq k\geq0$. This means, $k\neq m-M +\sum_{i=1}^a n_i=i(a), \ N-1\geq k\geq0$. Therefore, $i(a)>N-1$ which means $i(a)\geq N$. By equation (\ref{genfusion}), one has $P^{(m,\underline n)}*Q^{(M,N)}=0$. Here $P^{(m,n)}$ is the defect with the set $(m,n)$  a solution to $\mathcal{L}^c_{(m,n)}=\left\{ M,\dots, M+N-1\right\}$; and $Q^{(M,N)}$ such that $M= \min\left\{ M,\dots, M+N-1\right\}$, and $N =|\left\{ M,\dots, M+N-1\right\}|$.

Now if $\mathcal{S}_{\mathcal{L}^c_{(m,n)}}$ does not delete the full union  $S^{(M,N)}$, then
\begin{equation}\label{diskintersection}
\begin{split}
\left\{ M,\dots, M+N-1\right\}\cap \mathcal{L}^c_{(m,n)}&=\left\{ 0,\dots, N-1\right\}\cap m -M+\left\{n_0,n_0+n_1,\dots, n_0+n_1+\cdots +n_{d'-1}\right\}\\
&=\left\{ 0,\dots, N-1\right\}\cap \mathcal{J}\\
&=\left\{i_1,\dots,i_l\right\}\neq \emptyset,
\end{split}
\end{equation}
where $\mathcal{J}:=\left\{i(a)\ |\ a\in \mathbb{Z}_{d'}\right\}$. Hence, there exists $a\in \mathbb{Z}_{d'}$ such that $i(a)< N$ and by equation (\ref{genfusion}) the corresponding fusion $P^{(m,n)}*Q^{(M,N)}$ is not zero. As previously discussed this fusion is then $P^{(m,n)}*Q^{(M,N)}=Q^{(a_1,k(a_1)}$ where $a_1$ minimizes $i(a)$ and
\begin{equation}
k(a)=\min\left\{j>0 \ | \ \sum_{l=1}^j n_{a+l} \leq N\right\}.
\end{equation}
Since we have restricted to the case $n_i\geq 1 \ \forall i$, $k(a_1)=l$ is the number of sections of $S^{(M,N)}$ not annihilated by $S_{\left\{\cdots\right\}}$. Thus, $P^{(m,\underline n)}*Q^{(M,N)}=Q^{(a_1,l)}$. {One notes that $a_1$ is the number of $Q^{(M',1)}$ with $M'\in\left\{m,\dots, M\right\}$ not annihilated by $P$.} Hence, the operators $\mathcal{O}$ represent the $P$  action on all B-type boundary conditions \cite{brunner07a}.

\subsection{Comparison with RG flow}\label{comparison}

The RG flows between the $\mathbb{C}/\mathbb{Z}_d$ orbifolds can be studied in terms of the mirror picture as well. As we previously mentioned, mirror symmetry relates these orbifolds and the twisted Landau-Ginzburg model with twisted superpotential $\widetilde W = \widetilde X^d$. These twisted model can be related via mirror symmetry to a Landau-Ginzburg model with superpotential $ W =  X^d$. Therefore we can frame the RG flow of interest $\mathbb{C}/\mathbb{Z}_d\longrightarrow \mathbb{C}/\mathbb{Z}_{d'}$ as the RG flow $\text{LG}_d \longrightarrow \text{LG}_{d'}$ in the presence of A-supersymmetry.

The RG flows in the Landau-Ginzburg models are encoded in the behavior of the deformed superpotential $W_\lambda$ of the respective model. That is, we consider perturbations 
\begin{equation}
W_{\lambda_0}=X^d+\lambda_0 X^{d'}\ \ , \ \ d'<d,
\end{equation}
of $W=X^d$. The RG flow affects the superpotential by scaling it
\begin{equation}\label{rgscale}
W_{\lambda_0}\rightarrow \Lambda^{-1} W_{\lambda_0}.
\end{equation}
Upon a field redefinition, $X\rightarrow \Lambda X$, we obtain
\begin{equation}\label{deformedW}
\Lambda^{-1}W_{\lambda_0}=X^d+\lambda_0\Lambda^{\frac{d'-d}{d}}X^{d'}=:W_\lambda(X),
\end{equation}
where $\lambda(\Lambda):=\lambda_0\Lambda^{\frac{d'-d}{d}}$ is the running parameter:
\begin{equation}
\lim_{\Lambda \rightarrow \infty} \lambda = 0\ \ (\textit{UV}) \ , \ \ \lim_{\Lambda \rightarrow 0} \lambda = \infty \ \ (\textit{IR}).
\end{equation}
So at either end of the flow we end up with a homogeneous potential. We assume that the imaginary parts of the critical values of $W_\lambda$ stay different $\forall \ \lambda$.

Since we are interested in Landau-Ginzburg models on the half-plane with a non-zero boundary, we refer to the language of A-branes discussed in Section \ref{abraneslg}. The RG flow has a description in terms of the A-branes and the respective deformations \cite{hori00, brunner07} under non-zero $\lambda$ in equation (\ref{deformedW}). Each A-brane formed by segments from $X_{*i}$ to the boundary points $B_a$ and $B_b$ is denoted by $\overline{B_a X_{*i} B_b}$. As the deformed superpotential flows into the IR, the critical points $X_{*i}$, $i>0$, flow to infinity, while the critical point $X_{*1}=0$ associated with the homogeneous superpotential remains. The A-branes associated with the points $X_{*i}$ then decouple from the theory since the respective Lagrangian submanifolds $\gamma_{X_{*i}}$ disappear.  Therefore the IR A-branes are labeled by the equivalent classes $([B_i],[B_j])$ of the relationship $B_k\sim B_l$ when connected on $\Gamma\setminus\Gamma_1$. A generic A-brane in the UV might be composed of segments which are part of $\Gamma_1$ and $\Gamma_i$ in the deformed potential $(\lambda\neq0)$. In this case the A-brane decays into the sum of an A-brane which decouples in the IR and an A-brane which flows to an IR A-brane.

To illustrate, let us consider the example we discussed in Section \ref{abraneslg} with $W=X^4$ and the deformation $W_\lambda=X^4+\lambda X^3$. $W=X^4$ corresponds to the $\mathbb{C}/\mathbb{Z}_4$ orbifold. The deformed $W_\lambda$ has critical points $X_{*1}=0$ of order $n=2$, and $X_{*2}=-3\lambda$ of order $n=1$. We see that we flow to the IR $X_{*2}\rightarrow \partial D$ so the A-brane $\overline{B_3 X_{*2}B_2}$ decouples. So the endpoint of the flow is the $\mathbb{C}/\mathbb{Z}_3$ orbifold. As an example of the decay of the UV A-branes when $\lambda\neq 0$, consider $\overline{B_3 X_{*}B_1}$. As we turn on $\lambda$ this A-brane decays to $\overline{B_3 X_{*2}B_2} +\overline{B_2 X_{*1}B_1}$.

One can map the A-brane diagrams to the disk diagrams representing the B-type boundary conditions \cite{brunner07}; and hence there is a correspondence between the flow of the A-brane deformations and the action of the special defects on the disk diagrams of B-type boundary conditions. As noted above, in the IR only those preimages of $\infty$ which are not connected on $\Gamma \setminus\Gamma_1$ survive. These are precisely the points in the set
\begin{equation}\label{pointset}
\mathcal{L}=\left\{a\in\mathbb{Z}_d | B_a \nsim  B_{a+1}\right\}.
\end{equation}
In terms of the graphical disk operations for the B-type defects, this is equivalent to starting with disk partitioned into $S_i$ sectors representing the $Q^{M,N}$ B-type boundary conditions; and acting on this disk with the $\mathcal{S}_{\mathcal{L}^c}$ operator with $\mathcal{L}$ as in  equation (\ref{pointset}).

\section{Summary and outlook}

In this note we have presented an example of topological defects which implement the action of the RG flow between $\mathbb{C}/\mathbb{Z}_n$ 
theories.  The language we have employed to describe the RG flow defects is the natural description for such objects in the frame of Landau-Ginzburg models and their orbifolds. As we reviewed in Section \ref{btypeDefects}, this description involves factorizing the superpotentials of the given theories over different polynomial rings. 

Here we have showed that the language of matrix factorizations for boundaries and defects carries over to the case of a zero superpotential. The matrix factorizations we used in this case were obtained by setting $p_0=0$ in those given in \cite{brunner07a}. This is a very natural choice since it relates matrix factorizations in the $\mathbb{C}/\mathbb{Z}_d$ models to another method of characterizing D-branes.  Indeed, a common description of D-branes in geometric spaces (when there is no superpotential) is via chain complexes of vector bundles, with a differential $d$ built from the BRST operator $Q$ \cite{aspinwall}. On the other hand, out of the matrix factorizations associated with the D-branes in the Landau-Ginzburg models one obtains 2-periodic twisted complexes by taking the differentials to be the factorizing maps $p_1$ and $p_0$. Therefore with $p_0=0$,  $W\rightarrow 0$ produces an ordinary complex which coincides with above description for the D-branes.  It would be interesting to make this connection precise in a more general context\footnote{We thank Ilka Brunner for emphasizing this connection to us.}.

We have put forward two different ways of checking that the defects we posit in this note indeed enforce the RG flow between the non-compact orbifolds. One method uses the chiral rings of the theories at hand, and their deformations. The other method is a geometrical description of A-branes which are the equivalent representation of B-type boundary conditions in the mirror theory. Both methods keep track of the RG flow and show that the endpoints  are $\mathbb{C}/\mathbb{Z}_n$ orbifolds. The defects $P^{(m,\underline n)}$ of Subsection \ref{rgflowDefects} are shown to be appropriate interfaces between any two such orbifolds. 

By studying the fusion rules we showed that we can use these defects to tackle the question of the boundary RG flow when the theory has a nontrivial worldsheet boundary. In this note we provided evidence that the defects $P^{(m,\underline n)}$ successfully map the boundary conditions associated with the IR theory $\mathbb{C}/\mathbb{Z}_n$, to those of the UV theory $\mathbb{C}/\mathbb{Z}_n'$, $n'<n$. We established such correspondence by working with the mirror theory of the non-compact orbifolds. In this picture, we can compare the action of the RG flow defects on the B-type D-branes with the action of the RG flow on the dual A-type D-branes, i.e., A-branes. In comparing with the work of \cite{brunner07a}, we have shown that the RG flows between the $\mathbb{C}/\mathbb{Z}_d$ models follow a similar pattern to that of the LG orbifolds with a superpotential turned on.

Although we checked that RG flow defects properly describe the bulk-induced boundary RG flow by going to the mirror description in Subsection \ref{comparison}, a  similar comparison can be done between the result of the fusion rules and the flow of the deformed relation of the chiral ring given in equation (\ref{deformed}).  This can be done by considering the quotient relation of the chiral ring in equation (\ref{deformed}) as a branched covering of the complex plane. Such a description would provide an equivalent geometrical formalism to that of the deformed A-branes, so that an analysis could be done along the lines of the one done in Subsection \ref{comparison} for the A-branes.

A different approach to building conformal defects in these non-compact orbifolds is via the unfolding procedure described in \cite{bachas07}. In this method one constructs the boundary states corresponding to D-branes in the target space  $\mathbb{C}/\mathbb{Z}_n \times \mathbb{C}/\mathbb{Z}_{n'}$. These states can be mapped to defects between the theories $\mathbb{C}/\mathbb{Z}_{n}$ and $\mathbb{C}/\mathbb{Z}_{n'}$ via the inverse process of the ``folding trick''. An interesting question would be to find an equivalent description of the RG flow defects presented here in terms of the unfolding prescription. 

\section*{Acknowledgements}

We thank Andrew Royston and Ilka Brunner for comments on the manuscript. DR would like to thank Sebastian Guttenberg for early discussions. This work was supported by NSF under grants PHY-121433 and PHY-1521099 and the Mitchell Institute for Physics and Astronomy.

\bibliographystyle{utphys}

\bibliography{BibliographyDefectsLandauGinzburg}

\end{document}